\documentclass[pre,preprintnumbers,amsmath,amssymb]{revtex4-1}
\usepackage{graphicx,color}
\usepackage{bm}
\usepackage{epsfig}
\providecommand{\mre}[1]{\textcolor{black}{#1}}

\begin{document}
\title{A  sluggish  random walk with subdiffusive spread}
\author{Aniket Zodage$^{1,2}$, Rosalind J. Allen$^{3,4}$, Martin R. Evans$^{4,5}$, Satya N. Majumdar$^{5}$}
\vspace*{5ex}

\address{$^1$ Department of Physics, Indian Institute of Science Education and Research, Dr. Homi Bhabha Road, Pune 411008, India}
\address{$^2$ Department of Physics, UC San Diego, 9500 Gilman Dr. La Jolla, California 92093, USA }
\address{$^3$ Theoretical Microbial Ecology, Institute of Microbiology, Faculty of Biological Sciences, Friedrich Schiller University Jena, Buchaer Strasse 6, 07745 Jena, Germany}
\address{$^4$ SUPA, School of Physics and Astronomy, University of Edinburgh, Peter Guthrie Tait Road, EH9 3FD}
\address{$^5$ LPTMS, CNRS, Univ.~Paris-Sud, Universit\'e Paris-Saclay, 91405 Orsay, France}

\date{\today}
\begin{abstract}
 We study a one-dimensional sluggish random walk with space-dependent transition probabilities between nearest-neighbour lattice sites. Motivated by trap models of slow dynamics, we consider a model in which the trap depth  increases logarithmically with distance from the origin. This leads to a random walk which has symmetric transition probabilities that decrease  with distance $|k|$  from the origin as $1/|k|$ for large $|k|$.
We show that the typical position after time $t$ scales as $t^{1/3}$ with a nontrivial scaling function for the position distribution which has a trough (a cusp singularity) at the origin. \mre{Therefore an effective central bias away from the origin  emerges} even though the transition probabilities are symmetric. We also compute the survival probability of the walker in the presence of a sink at the origin and show that it decays as $t^{-1/3}$ at late times. Furthermore we compute the distribution of the maximum position, $M(t)$, to the right  of the origin up to time $t$,
and show that it has a nontrivial scaling function. Finally we provide a generalisation of this model where the transition probabilities decay as $1/|k|^\alpha$ with $\alpha >0$. 
\end{abstract}
\maketitle
\section {Introduction}
Slow dynamics is a common feature of many physical systems, including glasses, granular media and colloids \cite{book,book2}.
Slow dynamics commonly arises when the system becomes trapped
for increasing periods of time in deeper and deeper local free energy minima in the configuration space. 
This phenomenon has inspired the study of simplified toy models known as  trap models \cite{Bouchaud92,BD95,MB96,BB03,Sollich03}. In these models, the many minima of the complex disordered landscape are represented by traps whose  depths are taken to be random variables.
For a single particle hopping between nearby traps
the mean squared displacement typically grows more slowly than linearly in time,
thus  the particle's motion is subdiffusive \cite{BG90,MK00,BB05}.

Similar slow dynamics can also arise in an inhomogeneous, \mre{but non-random}, landscape where the trap depth is position-dependent
Here, the hopping dynamics between the traps is an example of a Markov chain with  space-dependent transition probabilities \cite{Hughes, MPW16}, or in other words, an inhomogeneous random walk. For such systems, explicit solutions  for observables beyond the simple position distribution  -- such as first passage probabilities \cite{Redner01,BMS13,MRO14} or extreme value statistics \cite{Gumbel,MPS20} -- are generally hard to obtain. 

A classic example of an inhomogenous random walk is the centrally-biased Gillis model \cite{Gillis56,OPRA20,PROA20,ROAP20,AOPR22}. In this model, a single particle hops on a one-dimensional lattice where the hopping probability is asymmetric in a position-dependent manner. Specifically, for $k\neq 0$, the hopping probabilities from site $k$ to $k\pm 1$ are $\frac{1}{2}(1\mp \epsilon/k)$, while for $k=0$ the hopping probabilities to sites $\pm 1$ are both $1/2$.  Because the hopping probabilities (for $k\neq 0$) are asymmetric, the particle undergoes a biased random walk in which the parameter  $\epsilon \in [-1,1]$ controls the strength of the bias. For $\epsilon >0$ there is  a drift towards the origin while for $\epsilon <0$ there is a drift away from the origin. Far away from the origin, where $|k| \gg 1$, the bias is small and the dynamics tends towards a symmetric random walk which, in the continuum limit, reduces to a particle moving in a logarithmic potential $U(k) \to  2 \epsilon \ln |k|$ \cite{OPRA20}.

The Gillis model  \cite{Gillis56,OPRA20,PROA20,ROAP20,AOPR22}, and its continuous limit of particle motion in a logarithmic potential \cite{Bray00,DLBK11,HMS11,LMS05,RR20,OPRA20}, 
 have aroused much  interest because of their  relevance to vortex dynamics, interactions between tracer particles in a driven fluid,  cold atoms trapped in optical lattices and the nonequilibrium behaviour of systems with long-range  interactions \cite{CDCT91,MEZ96,Lutz04,BD05,CDR09}.  These models have the appealing feature of allowing the exact calculation of various observables going beyond the position distribution.

Motivated by these works, here we consider the counterpart problem of an inhomogeneous trap model in which the trap depth increases  logarithmically  with increasing distance from the origin. The dynamics of a particle in this model  corresponds to a symmetric random walk with space-dependent hopping probabilities, \mre{between nearest-neighbour lattice sites,} that decrease
inversely with distance  from the origin. This  random walk has the interesting property of being  `sluggish'
since the particle's motion slows down as it goes further away from the origin.
We show that the physics of this model is quite different from the previously studied case of a particle in a  logarithmic potential. Instead, in the continuum limit and for $|k| \gg 1$, our model corresponds to a particle moving in a potential $U(k) \sim 1/|k|$ with, additionally, a
space-dependent diffusion constant that also decays as $1/|k|$. The interplay of these two features leads to an emergent bias in the dynamics away from the origin, even though the hopping probabilities are symmetric. The position distribution has a non-trivial and non-Gaussian form in  which distance scales with time as $t^{1/3}$ at late times.   Moreover, we show that other
observables such as  the survival probability  in the presence of an absorbing site and the distribution of the maximum displacement to one side of the origin can be  computed explicitly and exhibit
non-trivial scaling behaviour.  Finally we discuss how the model can be easily generalised to higher
dimensions and other space-dependent hopping probabilities, such as a $1/|k|^\alpha$ decay with exponent $\alpha >0$, without losing its solvability.

\begin{figure}[h!]
\includegraphics[width=1.0\textwidth]{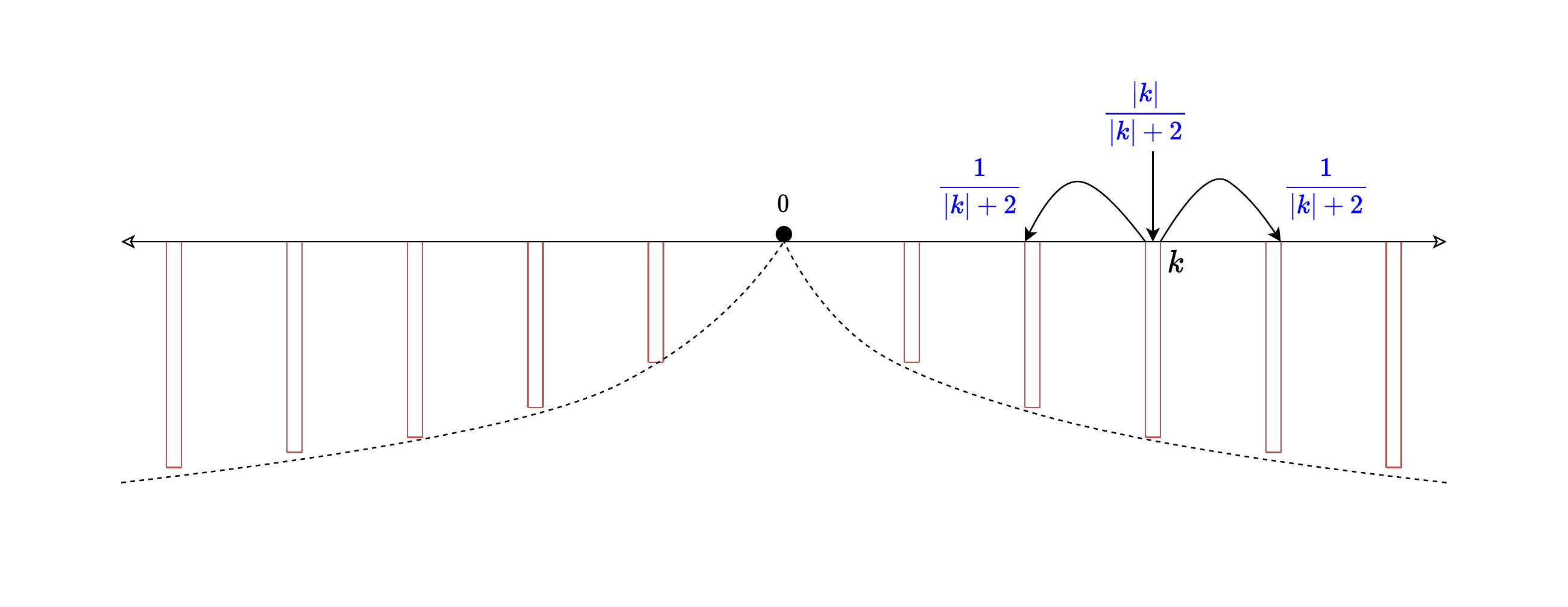}
\caption{Schematic illustration of our model, showing the ordered arrangement of traps and the  hopping probabilities $1/(|k|+2)$ for a particle to move  to a nearest neighbour of site  $k$ on the lattice, as well as the probability  $|k|/(|k|+2)$  of remaining at site $k$. }
\label{fig.trap}
\end{figure}

\section{Model definition and continuum limit}\label{sec:def}
As discussed, we consider an ordered array of traps arranged on a one-dimensional lattice such that the depth of the trap at site $k$ is
$a\ln(|k| + 2)$ (as illustrated in Fig.~\ref{fig.trap}). The corresponding Arrhenius escape rate from the trap at site $k$ is
$A (|k| +2)^{-\alpha}$ with $\alpha = \beta\, a$,
where $A$ is an overall constant and $\beta$ is the inverse temperature.
Without loss of generality we will set $A=1$.
We will mostly focus on the case where $\alpha=1$, although we briefly 
discuss $\alpha \neq 1$
in Section \ref{sec:alphannot1}. 

We consider the discrete-time dynamics of 
a particle moving at random on this infinite one-dimensional lattice. The key feature of the dynamics is that, as time progresses,  the particle explores sites further and further away from the origin in which it gets trapped to a greater and greater extent because of the increasing trap depths. Hence the particle is subject to diminishing  transition probability for exiting the traps.

At each integer time step $t$ the particle's position  evolves according
to the following rules (illustrated in Fig.~\ref{fig.trap}). From a site $k$ at time $t$, the particle hops to site $k+1$ with probability $1/(|k|+2)$,
it hops to  site $k-1$ with equal probability $1/(|k|+2)$, or it stays at site $k$ with the complementary probability $|k|/(|k|+2)$. The time is then updated to $t+1$. We note that (in contrast to the Gillis model \cite{Gillis56,OPRA20,PROA20,ROAP20,AOPR22}) the hopping probabilities are symmetric for all $k$; however they differ from those of a simple random walk everywhere except at the origin, $k=0$, where  the hopping probability is $1/2$ to either of $k= \pm 1$.

Let $P(k,t)$ denote the position distribution at time $t$ for a particle that starts from $k_0=0$ at $t=0$. The distribution  evolves
 via the forward master equation:
\begin{equation}
P(k,t+1)= \frac{1}{|k+1|+2}\, P(k+1,t) + \frac{1}{|k-1|+2}\, P(k-1,t) + \frac{|k|}{|k|+2}\, P(k,t)\, ,
\label{ffp.1}
\end{equation}
where the initial condition is $P(k,0)= \delta_{k,0}$.
The solution $P(k,t)$ is symmetric around $k=0$, hence we
can just focus on $k\ge 0$.  We first write \eqref{ffp.1} 
in the more suggestive form
\begin{equation}
  P(k,t+1)-P(k,t)=
\frac{1}{|k+1|+2}\, P(k+1,t) + \frac{1}{|k-1|+2}\, P(k-1,t) - \frac{2}{|k|+2}\, P(k,t)\, .
\label{ffp.2}
\end{equation}
For large $k>0$ and large $t$, we can expand the right hand side (rhs) of 
equation (\ref{ffp.2}) as a Taylor series in $k$ and replace the 
left hand side (lhs) by a time derivative.
This gives, keeping all terms of the same order, the \mre{continuum}
equation that captures the behaviour of the system at long distance and at late time:
\mre{\begin{eqnarray}
\frac{\partial }{\partial t}P(k,t)&\approx& \frac{\partial^2}{\partial k^2} \left[ \frac{1}{k} P(k,t) \right]  \label{diffeq}\\[1ex]
&=& \frac{1}{k}\left[ \frac{\partial^2}{\partial k^2} P(k,t) - \frac{2}{k}\, \frac{\partial }{\partial k} P(k,t)
+\frac{2}{k^2}\, P(k,t)\right]\, .
\label{ffp_scaling.1}
\end{eqnarray}}
Thus, the continuum limit of the sluggish random walk yields a diffusion equation of the form  \eqref{diffeq}  with space-dependent diffusion constant $D(k) = 1/k$.

This  equation can be written as a continuity equation:
\begin{equation}
\frac{\partial }{\partial t}P(k,t)= -\frac{\partial}{\partial k} j(k,t)\, ,
\label{cont.1}
\end{equation}
where the space-time dependent current density $j(k,t)$ reads
\begin{equation}
j(k,t)= -\frac{1}{k}\, \frac{\partial }{\partial k} P(k,t) + \frac{1}{k^2}\, P(k,t)\, .
\label{current.1}
\end{equation}
\mre{Now comparing with the more familiar Smoluchowski form of the diffusion equation for a Brownian particle with diffusion coefficient $D(k)$ in a potential $U(x)$
\begin{equation}
\frac{\partial }{\partial t} P(k,t) =\frac{\partial}{\partial k} \left [D(k) \frac{\partial}{\partial k} P(k,t)+
   \left(\frac{\partial}{\partial k} U(k)\right) P(k,t)\right]\, ,
  \label{ffp3}
\end{equation}
we identify the first term on the rhs of \eqref{current.1} as a diffusive probability  current with $D(k) = 1/k$  and the second term as an \mre{outwards} drift away from the origin due to an effective potential $U(k) = 1/k$.
Thus the Smoluchowski form, equation (\ref{ffp3}),  illustrates the features of the dynamics that lead to non-trivial behaviour. First, the  diffusion constant $D(k)=1/k$ is  space dependent, such that it slows down the dynamics as  the particle moves away from the origin. Additionally, an effective external potential $U(k)=1/k$ emerges, which is 
repulsive and pushes the particle away from the origin.} This repulsion arises
from the microscopic dynamics: 
the hopping probability
from site $k$ to $k+1$ is $1/(k+2)$, while the reverse event (from $(k+1)$ to $k$)
has probability $1/(k+3)< 1/(k+2)$ for any $k>0$. Thus, even though the hopping probabilities out of site $k$ (i.e. to ($k+1$) or ($k-1$)) are symmetric, the space-dependence of the hopping probability produces   an outward bias
away from the origin (symmetrically for $k<0$) which
leads to the \mre{outwards} drift term in the current in equation (\ref{current.1})
in the continuum description.

For large time and space we expect to see a scaling regime in which the 
probability distribution becomes a function of a combination  
$k/t^{\nu}$ (with $\nu>0$). The following argument suggests that $\nu$ takes the value $1/3$.  Let us assume that after time $t$, 
the typical value of the position $k$ is $k_{\rm typ}$. The number of steps $N$ that have been taken by the particle will scale as $N \sim t/k_{\rm typ}$, since the time for one step is the typical escape time $1/k_{\rm typ}$. Since all steps are equal in distance and the hopping probability is symmetric, the position scales with the number of steps in the same way as for a simple random walk, $k_{\rm typ} \sim N^{1/2}$. Putting these arguments together we obtain $k_{\rm typ} \sim N^{1/2} \sim ( t/k_{\rm typ})^{1/2}$ which implies that the position of the particle scales with time as
$k_{\rm typ} \sim t^{1/3}$; hence $\nu=1/3$.

This scaling can be confirmed by assuming the following scaling form for the probability distribution $P(k,t)$  in the 
 limit when both $k$ and $t$ are large, keeping the
ratio $k/t^{\nu}$ (with $\nu>0$) fixed:
\begin{equation}
P(k,t) \to \frac{1}{b\, t^{\nu}} G\left( \frac{k}{b\, t^{\nu}}\right)\, ,
\label{scaling_form.0}
\end{equation}
where $G(z)$ is the scaling function. We have also
incorporated an adjustable constant $b$ which can
be chosen appropriately. 
Substituting the scaling form (\ref{scaling_form.0}) in 
equation (\ref{ffp_scaling.1}), one readily finds that for leading order terms to be of 
the same order 
we must have 
$\nu= \frac{1}{3}$. For convenience we will also choose $b = 3^{1/3}$.
We will discuss the precise form of the scaling function $G(z)$
in Section \ref{sec:scaling}. 

The scaling $k\sim t^{1/3}$ also manifests itself 
in other observables, such as the survival probability and the
distribution of the maximum position of the random walk that we study in this paper.
In the next section, for clarity, we  summarize our main results. Then, in the following sections, we  
discuss each result in detail.

\section{Summary of key results}
In this paper we derive exact results in the scaling limit for three observables: the position distribution, the survival probability and the distribution of the maximum of the random walk. For clarity, we state these results here; their  derivations will be presented in the following sections.

\begin{description}
\item[Position distribution]
in the large $t$ and large $k$ limit, such that
$z =k/(3t)^{1/3}$ is fixed, the position distribution of the walker  $P(k,t)$ is given by
\begin{equation}\label{eq:pkt}
P(k,t) \to  \frac{1}{(3t)^{1/3}}\, G\left(\frac{k}{(3t)^{1/3}}\right)\,  
\end{equation}
where the scaling function $G(z)$ is given by
\begin{equation}\label{eq:gofz}
G(z)= \frac{3^{1/3}}{2\,\Gamma(2/3)}\, |z|\, e^{-|z|^3/3}\, .
\end{equation}
This function has a trough at $z=0$ and the function is bimodal with peaks at $z=\pm1$ (see Fig. \ref{fig.Gz}).

\item[Survival probability]  For a walker that starts from $k_0 >0$, the probability that the trap at $k=0$ has not been visited by time $t$ is equivalent to  the survival probability $Q(k_0,t)$ in the presence of an absorbing site at the origin $k=0$. This is given in the scaling limit by
\begin{equation}
Q(k_0,t) \approx f\left(\frac{k_0}{(3 t)^{1/3}}\right)\,, \quad {\rm where}\quad f(z)= 1- \frac{1}{\Gamma(1/3)}\,
\Gamma(1/3, z^3/3)\, ,
\end{equation}
(see Fig. \ref{fig.fz}).
This implies that in the long time limit the survival probability decays as $t^{-1/3}$
(see equation \eqref{Qsmallz}).
We also compute the joint distribution of position and survival (equations
\eqref{scaling_s.2} and \eqref{hz_sol.2}).

\item[Distribution of maximum]
  The distribution of $M$, the furthest site to the right visited by the walker up to time $t$, or equivalently the deepest trap visited up to time $t$, is given in the scaling limit by
\begin{equation}
  P(M=L,t) \to \frac{1}{(3t)^{1/3}}g\left( \frac{L}{(3t)^{1/3}}\right) \end{equation}
where $y=L/t^{1/3}$ is now the scaling variable. The scaling function
 $g(y)$ (see Fig. \ref{fig.gz})  is described in equations \eqref{PMLg},\eqref{glarge} and \eqref{gsmall}.
\end{description}

\section{Scaling form of the position distribution}\label{sec:scaling}

\begin{figure}[h!]
\includegraphics[width=0.8\textwidth]{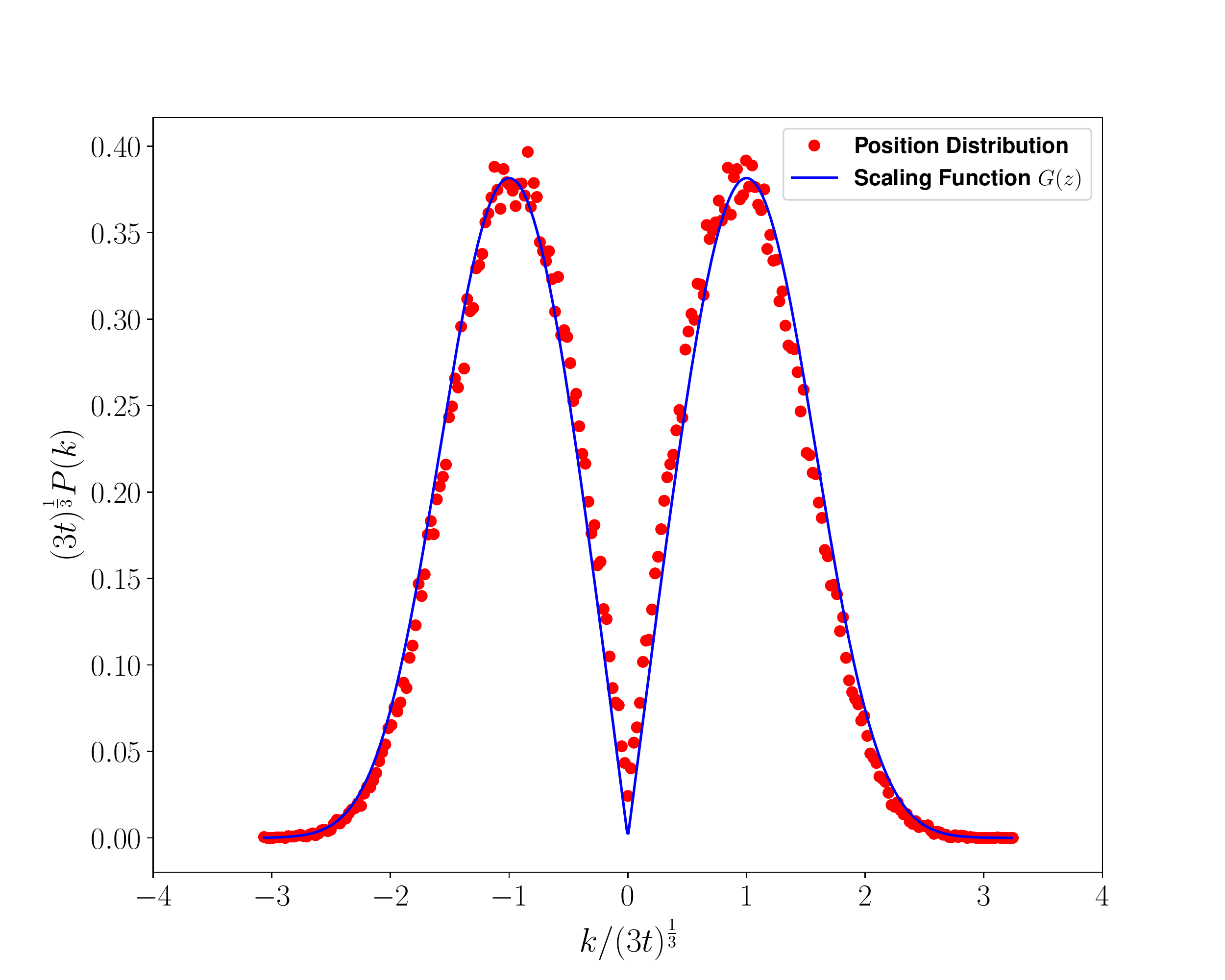}
\caption{The scaling function $G(z)$ plotted as a function of $z$. Symbols are obtained from Monte Carlo simulation data for the random walk. Starting from $k_{0} = 0$ at $t=0$, the random walk is numerically evolved up to $t=20000$. The symbols show the scaled  histogram of final positions obtained from $n=150000$ runs of the random walk simulation.}
\label{fig.Gz}
\end{figure}

We now derive equations \eqref{eq:pkt} and \eqref{eq:gofz} for the position distribution $P(k,t)$ of the random walk.
As discussed earlier, the form of  equation  (\ref{ffp_scaling.1}) implies that the correct scaling variable involving $k$ and $t$ is $z = k/(b t^{1/3})$, and we choose the arbitrary constant as $b = 3^{1/3}$ for later convenience.
Therefore, to solve the continuum equation (\ref{ffp_scaling.1}),
we  assume a scaling solution at late times and large $k$ of the
form 
\begin{equation}
P(k,t) = \frac{1}{(3t)^{1/3}}\, G\left(\frac{k}{(3t)^{1/3}}\right)\, , 
\label{Gz.1}
\end{equation}
where $G(z)$ is symmetric around $z=0$, and is normalized to $1$, i.e., 
$\int_{-\infty}^{\infty} G(z)\, dz= 1$, or equivalently
\begin{equation}
\int_0^{\infty} G(z)\, dz=\frac{1}{2}\, .
\label{Gz_norm}
\end{equation}
Substituting the scaling ansatz (\ref{Gz.1}) in equation (\ref{ffp.1}) and taking the
scaling limit $k\to \infty$, $t\to \infty$ keeping
$z= k/(3  t)^{1/3}$ fixed, we find that $G(z)$, for $z>0$,
satisfies a second order ordinary differential equation
\begin{equation}
G''(z)+ \left(z^2- \frac{2}{z}\right)\, G'(z)+ \left(z+ \frac{2}{z^2}\right)\, G(z)=0\, .
\label{diff_G}
\end{equation}
Remarkably,  the general solution of this differential equation can be expressed
in a simple closed form
\begin{equation}
G(z)= c_1\, z\, {\rm e}^{-z^3/3} + c_2\, z\, {\rm e}^{-z^3/3}\, \int_0^{z} {\rm e}^{u^3/3}\, {\rm d} u\, ,
\label{Gz_sol.1}
\end{equation}
where $c_1$ and $c_2$ are arbitrary. However, the second solution (the second term in \eqref{Gz_sol.1})  behaves,
for large $z$, as $1/z$, and hence is not normalisable, implying that we must have
$c_2=0$. The constant $c_1$ can be fixed via the normalization constant
$\int_0^{\infty} G(z) dz=1/2$. Using the symmetry around $z=0$, the full solution for the scaling distribution \eqref{Gz.1}
is then given by
\begin{equation}
G(z)= \frac{3^{1/3}}{2\,\Gamma(2/3)}\, |z|\, e^{-|z|^3/3}\, .
\label{Gz_sol}
\end{equation}

This function is plotted in Fig. (\ref{fig.Gz})  where we also plot results of Monte Carlo simulations that approach the scaling curve. Strikingly,
in contrast to a simple random walk (where the scaling variable is $z= k/(2t)^{1/2}$  and the corresponding scaling function is Gaussian with 
a peak at $z=0$),  $G(z)$ has a trough at $z=0$ where the solution has a cusp
singularity. The origin of this trough can be traced back to
the drift term (away from the origin) in the current in equation (\ref{current.1}),
that leads to a depletion
of probability density near the origin at long times.
Thus by creating an emergent current away from the origin, the sluggish dynamics that is manifested in our model keeps 
the particle away from the origin and produces two peaks (i.e. bimodality) 
in the probability distribution; these peaks are located at  $z=\pm 1$ or equivalently $|k| = (3t)^{1/3}$. 
The distribution of the depth of the trap occupied at time $t$ also follows from equation (\ref{eq:pkt}) since the trap depth is $a\ln(|k| + 2)$.

\section{Survival Probability}
\label{sec:Q}
We now introduce a sink at the origin, such that if the random walker
arrives at $k=0$, it dies. We consider the survival probability of the walker in the presence of this absorbing site at $k=0$.
The `survival probability' $Q(k_0,t)$ denotes the probability that the walker is still alive after $t$ steps, given that it starts at site $k_0$ at time zero.
Clearly, $Q(k_0,t)$ is symmetric in $k_0$, so we will consider only $k_0\ge 0$,
implying that the walk is defined on the positive integers.

It is convenient to use the backward master equation for the survival probability:
\begin{equation}
Q(k_0,t+1)= \frac{1}{k_0+2}\, Q(k_0+1,t) + \frac{1}{k_0+2}\, Q(k_0-1,t) + \left(1- \frac{2}{k_0+2}\right)\, Q(k_0,t)\, ,
\label{bfp.1}
\end{equation}
for $k_0 \geq 1$.
This equation has a simple interpretation, corresponding to the events that may occur in the first step of the walk. 
In the first step,
the walker either hops from site $k_0$ (rightwards to $k_0+1$, or leftwards to $k_0-1$), or it stays at $k_0$. 
Then, starting from its position at time step $1$, it has to survive
a further $t$ steps. Summing these three possibilities for the first step leads to equation (\ref{bfp.1}), which needs to be solved
for $k_0\ge 1$ with the boundary conditions
\begin{eqnarray}
Q(k_0=0, t) &= & 0 \label{bc0} \\
Q(k_0\to \infty, t) &=& 1 \, . 
\label{bcinf} 
\end{eqnarray}
The first boundary condition corresponds to the fact that if the walker starts at the absorbing site $k_0=0$ it dies immediately. The second condition follows from the fact that if the walker
starts far away from the origin, it survives with probability $1$ as long as $t$ is finite.
In the limit of continuous time $t$ and space $k$ the backward equation becomes
\begin{equation}
\frac{ \partial}{\partial t} Q(k_0,t)= \frac{1}{k_0}\, 
\frac{ \partial^2}{\partial k_0^2}Q(k_0,t)\, .
\label{bfp3}
\end{equation}

It is convenient to 
use a scaling approach to quickly derive the large $t$ asymptotic behaviour
of the survival probability. We aim to solve equation (\ref{bfp3}) in the scaling limit introduced in section \ref{sec:scaling} when both $k_0$ and $t$ are large. Following the discussion in section \ref{sec:scaling} we expect that $Q(k_0,t)$ will satisfy a scaling form
\begin{equation}
Q(k_0,t) \to  f\left( \frac{k_0}{(3 t)^{1/3}}\right)\, ,
\label{scaling_form.1}
\end{equation}
where $f(z)$ is the scaling function.
We now substitute the scaling form (\ref{scaling_form.1}) in equation (\ref{bfp3}) and expand to leading order to obtain
the following second order ordinary
differential equation in $z\ge 0$ for the scaling function
\begin{equation}
f''(z) = - z^2\, f'(z)\, ,
\label{fz.1}
\end{equation}
subject to the two boundary conditions
\begin{equation}
f(z=0)=0 \quad {\rm and} \quad f(z\to \infty)=1\, ,
\label{fz_bc.1}
\end{equation}
which follow from Eqs. (\ref{bc0}) and (\ref{bcinf}) respectively.

\begin{figure}[h!]
\includegraphics[width=0.8\textwidth]{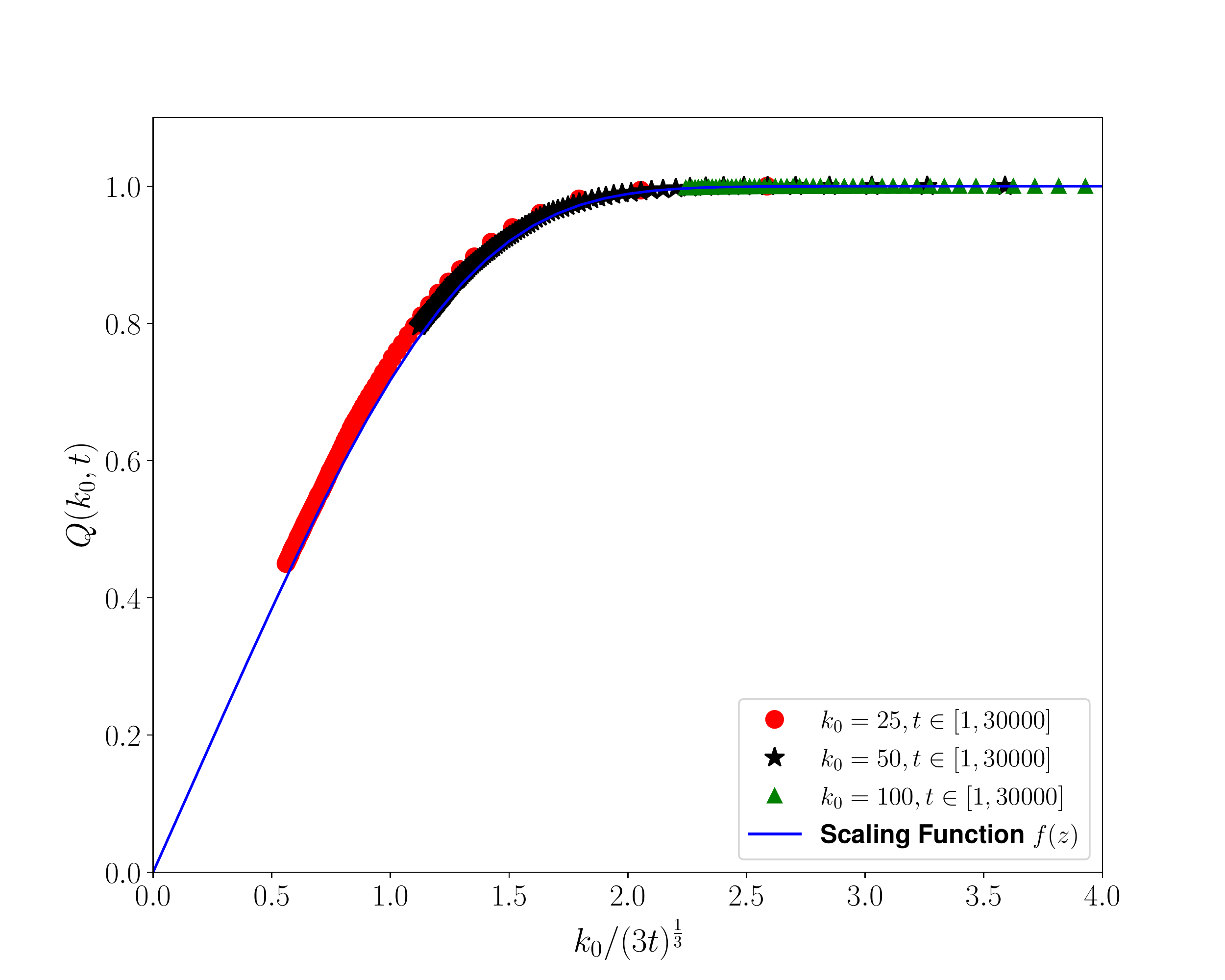}
\caption{Full curve: the scaling function $f(z)$ plotted as a function of scaling variable $z$. Symbols are obtained from Monte Carlo simulation data for different values of $k_0$.  For a given $k_{0}$, the random walk is numerically evolved over the time window shown in the legend. The symbols show histograms obtained from $n=10000$ runs of the random walk simulation. These histograms  are plotted against the scaling variable $z = k_{0}/(3t)^{1/3}$ .  The different intervals of $z$ for different values of $k_{0}$ are chosen for purposes of clarity.}
\label{fig.fz}
\end{figure}

The solution of equation (\ref{fz.1}) can be found trivially.
Integrating (\ref{fz.1}) once  gives $f'(z)= C\, {\rm e}^{-z^3/3}$. Integrating once more, using the boundary conditions (\ref{fz_bc.1}), leads
to the exact solution for the scaling function:
\begin{equation}
f(z)= \frac{\int_0^z {\rm e}^{-x^3/3}\, dx}{\int_0^{\infty} {\rm e}^{-x^3/3}\, {\rm d} x}= 1- \frac{1}{\Gamma(1/3)}\,
\Gamma(1/3, z^3/3)\, ,
\label{scaling_final}
\end{equation}
where $\Gamma(s,x)= \int_{x}^{\infty} {\rm e}^{-t}\, t^{s-1}\, {\rm d}t$ is the incomplete Gamma function.
\mre{The scaling function $f(z)$ is plotted in Fig. (\ref{fig.fz})  where we also plot results of Monte Carlo simulations that approach the scaling curve, for large $k_0$ and $t$.}
The scaling function is linear for small $z$ ($t^{1/3} \gg k_0$)
and saturates at $f =1$ for large $z$ ($t^{1/3} \ll k_0$). More precisely, the
 scaling function has the asymptotic behaviours
\begin{eqnarray}
f(z)\approx \begin{cases}
\frac{3^{2/3}}{\Gamma(1/3)}\, z + O(z^4)\quad {\rm as}\,\,\, z\to 0 \label{z0}\\
\\
1- \frac{3^{2/3}}{\Gamma(1/3)\, z^2}\, {\rm e}^{-z^3/3} \quad {\rm as} \,\,\, z\to \infty\;.  \label{zinf}
\end{cases}
\end{eqnarray}
In particular, for $z \to 0$
\begin{equation}
  Q(k_0,t) \simeq  \frac{3^{1/3}}{\Gamma(1/3)} \frac{k_0}{t^{1/3}}\;.
\label{Qsmallz}
\end{equation}
Equation \eqref{Qsmallz} implies that in the limit $t\to \infty$ the  asymptotic behaviour of the survival probability is $Q \sim t^{-1/3}$. The exponent $1/3$ is smaller than the value $1/2$ that is obtained for a simple diffusive process, implying that the decay is  slower than for simple diffusion. Again, the sluggish dynamics results in a significant difference in the dynamical properties, compared to those of a simple random walk.

\section{Joint survival and position distribution}\label{sec:joint}

Next we consider the probability $P_s(k,t|k_0)$ that a walker, starting at $k_0>0$ at time $t=0$, arrives at  $k$ at time $t$, having in the meantime avoided the sink at $k=0$. This is  the joint distribution of survival and position, with the  subscript $s$ in $P_s(k,t|k_0)$
denoting survival.

For this calculation we use the forward master equation, for $k>0$:
\begin{equation}
P_s(k,t+1|k_0)= \frac{1}{k+3}\, P_s(k+1,t|k_0) + \frac{1}{k+1}\, 
P_s(k-1,t|k_0) + \frac{k}{k+2}\, P_s(k,t|k_0)\, ,
\label{ffp_s.1}
\end{equation}
with the boundary condition $P_s(0,t|k_0)=0$ and the initial condition, $P_s(k,t=0|k_0)= \delta_{k,k_0}$.
When summed over $k=1,2,\cdots$, one should recover the survival probability of section~\ref{sec:Q}, namely
\begin{equation}
\sum_{k=1}^{\infty} P_s(k,t|k_0)= Q(k_0,t)\, .
\label{bfp_s.1}
\end{equation}

\begin{figure}[h!]
\includegraphics[width=0.6\textwidth]{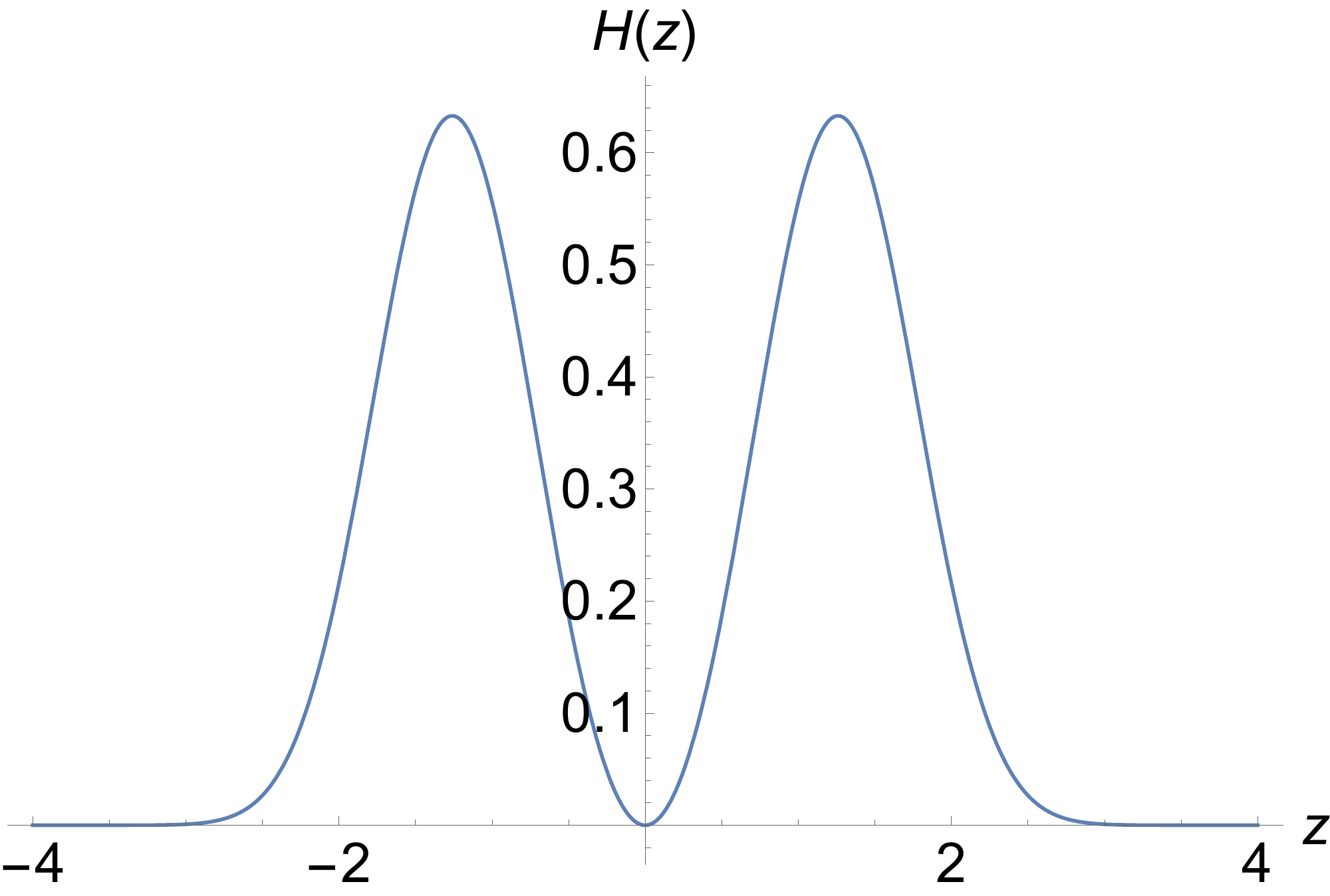}
\caption{The scaling function $H(z)$, given by equation \eqref{hz_sol.2}, plotted as a function of $z$.}
\label{fig.Hz}
\end{figure}

For simplicity, we will again work in the scaling limit where $t\to \infty$, $k\to \infty$ and $k_0\to \infty$, keeping
$z=k/(3 t)^{1/3}$ and $y=k_0/(3t)^{1/3}$ fixed. 
We expect a scaling form
\begin{equation}
P_s(k,t|k_0) \approx \frac{1}{(3t)^{1/3}}\, W\left(\frac{k}{(3t)^{1/3}}, \frac{k_0}{(3t)^{1/3}}\right)\, ,
\label{ffp_s.2}
\end{equation}
such that when integrated over $k$, we recover the scaling of the survival
probability survival probability $Q(k_0,t)$ in equation (\ref{scaling_final}) with
\begin{equation}
\int_0^{\infty} W(z,y)\, dz= f(y)\, ,
\label{scaling_s.1}
\end{equation}
where $f(y)$ is given in equation (\ref{scaling_final}). Here we assume $k_0\sim O(1)$, so that the second argument
of the scaling function $W$ in equation (\ref{ffp_s.2}) approaches zero. From 
the small argument behaviour of the survival probability in (\ref{Qsmallz}), we expect
that $W(z, y\to 0)\to y\, H(z)$. This leads us to the scaling ansatz, valid for any $k_0\sim O(1)$:
\begin{equation}
P_s(k,t|k_0)\approx \frac{k_0}{(3t)^{2/3}}\, H\left(\frac{k}{(3t)^{1/3}}\right)\, .
\label{scaling_s.2}
\end{equation}

Substituting this scaling ansatz in equation (\ref{ffp_s.1}), we get, to leading order in 
$1/t$, the following
ordinary differential equation for $H(z)$, for any $z\ge 0$ (for $z<0$, this function is symmetric, hence
we consider only $z\ge 0$):
\begin{equation}
H''(z)+ \left(z^2-\frac{2}{z}\right) H'(z) + \left(2 z+ \frac{2}{z^2}\right) H(z)=0\, .
\label{Hz.1}
\end{equation}
\mre{It is worth noting that this equation differs from the corresponding equation
  \eqref{diff_G} for the scaling function of the position distribution, only  through the
  factor of $2$ multiplying $z$ in the coefficient of $H(z)$.}
The scaling function $H(z)$ should satisfy the absorbing boundary condition $H(0)=0$. 
One more condition can be derived by substituting the scaling ansatz (\ref{scaling_s.2}) in
equation (\ref{bfp_s.1}), and taking the limit $y= k_0/(3t)^{1/3}\to 0$. Using the 
small $y$ behaviour of $f(y)$ in equation (\ref{z0}), we obtain  
the following condition:
\begin{equation}
\int_0^{\infty} H(z)\, dz= \frac{3^{2/3}}{\Gamma(1/3)}\, .
\label{cond_2.1}
\end{equation}
One can easily check that the normalised solution of (\ref{Hz.1}) is simply 
\begin{equation}
H(z)= \frac{3^{2/3}}{\Gamma(1/3)}\, z^2 e^{-z^3/3}\, .
\label{hz_sol.2}
\end{equation}
$H(z)$ is plotted in Fig. (\ref{fig.Hz}).
We note that the trough around $z=0$ is quadratic in $z$ for this calculation in the presence of a sink, in contrast to the linear $|z|$ dependence for the trough in the position distribution for the calculation without a sink (equation \eqref{Gz_sol}).
The quadratic behaviour of $H(z)$ near the origin also contrasts with the analogous result for the  simple random walk  case where linear behaviour is obtained as  $z\to 0$. This limit $z\to 0$ gives information on the long time behaviour; from  \eqref{scaling_s.2},\eqref{hz_sol.2} we obtain 
\begin{eqnarray}
  P_s(k,t|k_0)
   &\simeq & \frac{k_0 k^2}{3^{2/3} \Gamma(1/3)}   \, \frac{1}{t^{4/3}} \;.\label{Pka}
\end{eqnarray}
The $t^{-4/3}$ long-time behaviour of the survival probability in equation \eqref{Pka}  contrasts with the corresponding $t^{-3/2}$ behaviour for a simple random walk.

\section{Distribution of the maximum of the random walk}\label{sec:max}

We now remove the sink at the origin and instead consider a walker that starts at the origin  ($k_0=0$) and moves freely. We  study the statistics
of its maximum displacement $M(t)$ on the positive side up to time $t$.
This corresponds to the deepest trap visited to the right of the origin up to time $t$.
Then the cumulative distribution ${\rm Prob.}\left[M(t)\le L\right]$
is just the probability that the walker, starting at the origin, does
not visit the site $L$ up to time $t$.
Let $S(k_0,t)$ denote the probability that starting from $k_0$ at $t=0$,
the walker does not visit $L$ up to $t$. We then have 
\begin{equation}
{\rm Prob.}\left[M(t)\le L\right]= S(0,t)\, .
\label{cumul.1}
\end{equation}
To compute $S(0,t)$, we will first solve $S(k_0,t)$ for a general starting point
$k_0$ and then set $k_0=0$. The survival probability $S(k_0,t)$
again evolves according to the backward master equation
\begin{equation}
S(k_0,t+1)= \frac{1}{|k_0|+2}\, S(k_0+1,t) + \frac{1}{|k_0|+2}\, S(k_0-1,t) + \left(1- \frac{2}{|k_0|+2}\right)\, S(k_0,t)\, ,
\label{bfp.3}
\end{equation}
with boundary condition 
\begin{equation}
  S(L, t) =0\;, \label{QLbc}
\end{equation}
i.e. we impose a sink at site $k=L$.
The  initial condition (starting from $k_0<L$) is
\begin{equation}
  S(k_0,0) = 1\;.
  \end{equation}

Following the approach of section \ref{sec:Q},  we expand in $k_0$ to obtain the backward Fokker Planck  equation:
 \begin{equation}
\label{surv.1}
   \frac{\partial}{\partial t} S(k_0,t) = \frac{1}{|k_0|}
     \frac{\partial^2}{\partial k_0^2} S(k_0,t )\, ,
 \end{equation}
which is valid for $k_0\le L$, with an absorbing boundary condition $S(k_0=L,t)=0$ at the sink $k=L$  
and  the initial condition $S(k_0,0)=1$ for all $k_0<L$.

To solve equation (\ref{surv.1}), it is convenient to consider the Laplace transform
\begin{equation}
\widetilde{S}(k_0,s)= \int_0^{\infty} S(k_0,t)\, e^{-s\, t}\, dt\, .
\label{surv_lt.1}
\end{equation}
This satisfies
\begin{equation}
\frac{\partial^2}{\partial k_0^2} \widetilde S(k_0,s ) = s|k_0|  \widetilde 
S(k_0,s) -|k_0|\; ,
  \label{QLT1}
  \end{equation}
where we used the initial condition $S(k_0,0)=1$. Due to the presence
of the absolute value $k_0$ in the differential equation (\ref{QLT1}),
we need to solve for $0\le k_0\le L$ and $k_0\le 0$ separately,
and then match the solution and its first derivative at $k_0=0$.

The general solution of \eqref{QLT1} for $0\le k_0\le L$ and $k_0\le 0$ reads
  \begin{eqnarray}
    \widetilde S (k_0,s) &=& \frac{1}{s} + a_1\, \mbox{Ai} (s^{1/3}k_0)+ 
b_1\, \mbox{Bi} (s^{1/3}k_0)\quad\mbox{for}\quad 0\le k_0\le L  \label{kgt0}\\
        \widetilde S(k_0,s) &=& \frac{1}{s} + a_2\, \mbox{Ai} (-s^{1/3} k_0) 
\quad\quad\quad\quad\quad\quad \quad\mbox{for}\quad k_0\leq 0 \;, \label{klt0}
\end{eqnarray}
where $\mbox{Ai}(x)$ and $\mbox{Bi}(x)$ are the two linearly independent solutions
of the Airy differential equation $U''(x)- x U(x)=0$. Since $Bi(-x)$ 
diverges as $x\to-\infty$,
we discarded this in the solution for $k_0\le 0$ in equation (\ref{klt0}). 
The three constants (independent of $k_0$)  
$a_1$, $a_2$, $b_1$ are fixed by the continuity of 
$\widetilde S (k_0,s)$, the continuity of 
$\partial_{k_0} \widetilde S (k_0,s)$ at 
$k_0=0$ and the absorbing boundary condition $\widetilde S(k_0=L,s)=0$,
which yield three linear equations. These three constants can 
then be straightforwardly determined explicitly (we do not give the 
details here). If the walker starts at $k_0=0$ (for simplicity),
from equation (\ref{klt0}), we just need the
constant $a_2(s)$ since
\begin{equation}
  \widetilde S(0,s) = \frac{1}{s} + a_2(s)\, \mbox{Ai} (0)\, .
  \label{QL0}
  \end{equation}
It turns out that the expression of $a_2(s)$ is rather simple:
\begin{equation}
 a_2(s) = \frac{1}{2\pi\, \mbox{Ai(0)}\, \mbox{Ai}'(0)\, s\, 
\mbox{Bi}(s^{1/3}\, L)}= -\frac{\sqrt{3}}{s\, \mbox{Bi}(s^{1/3}\,L)}\, ,
\label{a2s}
\end{equation}
where we used $\mbox{Ai}(0)= 3^{-2/3}/\Gamma(2/3)$ and $\mbox{Ai}'(0)= -3^{-1/3}/\Gamma(1/3)$. 
Plugging in equation (\ref{QL0}) then gives the exact Laplace transform,
valid for all $s$:
\begin{equation}
\widetilde S(0,s)= \frac{1}{s}\left[1-\frac{1}{3^{1/6}\, \Gamma(2/3)}\,
\frac{1}{\mbox{Bi}(s^{1/3}\,L)}\right]\, .
\label{LT_3}
\end{equation}

\begin{figure}[h!]
\includegraphics[width=0.8\textwidth]{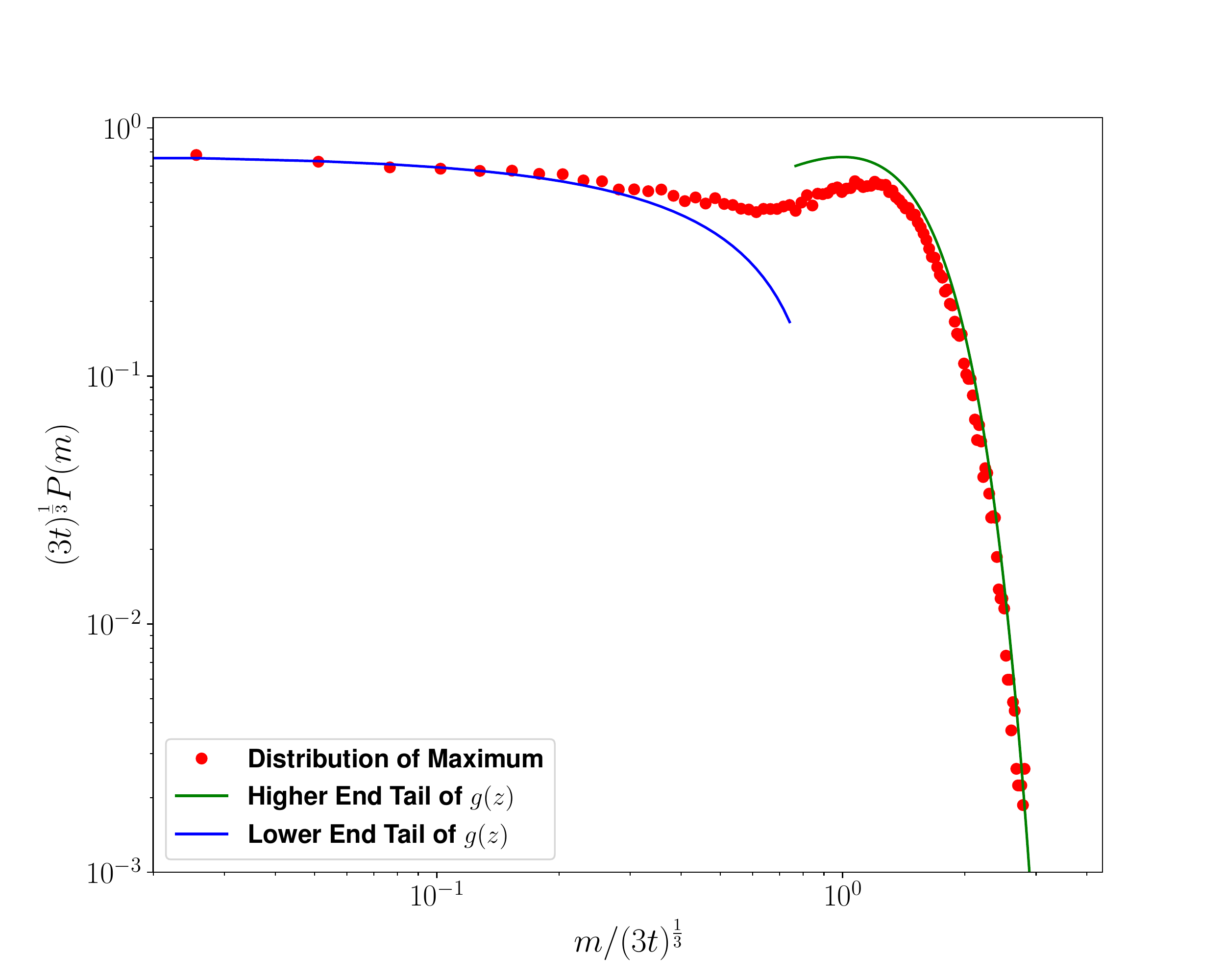}
\caption{Distribution of the maximum of the random walk. The two full curves denote the lower end tail of $g(z)$, equation \eqref{gsmall}, and higher end tail of $g(z)$, equation \eqref{glarge}. The symbols are obtained from Monte Carlo simulation data for the random walk. Starting from $k_{0}=0$ at $t=0$, the random walk was evolved up to $t=20000$. The symbols show the scaled  histogram obtained from $n=105000$ runs of the random walk simulation.}
\label{fig.gz}
\end{figure}

Taking the Laplace transform of equation (\ref{cumul.1}), and plugging in the
result (\ref{LT_3}), we obtain the exact Laplace transform of 
the cumulative distribution of the maximum:
\begin{equation}
 \int_0^{\infty} {\rm Prob.}\left[M(t)\le L\right]\, e^{-s\, t}\, {\rm d}t=
\frac{1}{s}\left[1-\frac{1}{3^{1/6}\, \Gamma(2/3)}\,
\frac{1}{\mbox{Bi}(s^{1/3}\,L)}\right]\, .
\label{cumul_lt.1}
\end{equation}
This result can be further simplified by noting that
${\rm Prob.}\left[M(t)\ge L\right]= 1- {\rm Prob.}\left[M(t)\le L\right]$.
Consequently,
\begin{equation}
 \int_0^{\infty} {\rm Prob.}\left[M(t)\ge L\right]\, e^{-s\, t}\,{\rm  d}t=
\frac{1}{3^{1/6}\, \Gamma(2/3)}\, \frac{1}{s\, \mbox{Bi}(s^{1/3}\,L)}\, .
\label{cumul_lt.2}
\end{equation}
Formally inverting this Laplace transform using the Bromwich contour
and rescaling $s L^{1/3}=\lambda$, one sees immediately that
for all $t$ and $L$, the cumulative distribution takes
the scaling form
\begin{equation}
{\rm Prob.}\left[M(t)\ge L\right]= Y\left(\frac{t}{L^{1/3}}\right)\, ,
\label{cumul_lt.3}
\end{equation}
where the scaling function $Y(y)$ has the exact Laplace transform
\begin{equation}
    \int_0^\infty {\rm e}^{-\lambda y}\, Y(y)\,{\rm d}y = 
\frac{3^{1/3}\Gamma(1/3)}{2\pi}
    \frac{1}{\lambda \mbox{Bi}(\lambda^{1/3})}\;.
        \label{Hlt}
    \end{equation}
While it is difficult to invert the Laplace transform exactly, 
it is straightforward to extract its asymptotic behaviours, as shown below.

The large $y$ behaviour of $Y(y)$ is controlled by the small $\lambda$ expansion  of \eqref{Hlt}
\begin{equation}
    \int_0^\infty {\rm e}^{-\lambda y}Y(y)\,{\rm d}y \simeq \frac{1}{\lambda} - \frac{
      3^{1/3}\Gamma(2/3) }{\Gamma(1/3)} \frac{1}{\lambda^{2/3}}
     + \frac{
      3^{2/3}\Gamma^2(2/3) }{\Gamma^2(1/3)} \frac{1}{\lambda^{1/3}} +\ldots\; ,
    \label{Hltsmall}
\end{equation}  
which yields the large $y$ asymptotic expansion
\begin{equation}
  Y(y) \sim 1- \frac{3^{1/3} }{\Gamma(1/3)} \frac{1}{y^{1/3}}  +\frac{3^{2/3} \Gamma^2(2/3) }{\Gamma^3(1/3)} \frac{1}{y^{2/3}} \ldots\;.
\end{equation}
The small $y$ behaviour of $Y(y)$ can be obtained from the large $\lambda$ asymptotic behaviour of \eqref{Hlt}
\begin{equation}
    \int_0^\infty {\rm e}^{-\lambda y}Y(y)\,{\rm d}y \sim \frac{ \pi^{1/2}}{3^{1/6} \Gamma(2/3) \lambda^{11/12} }\, {\rm e}^{-2/3 \ \lambda^{1/2}} \;,
    \label{Hltlarge}
\end{equation}
    which can be inverted to give the  small $y$ behaviour
\begin{equation}
Y(y) \simeq \frac{3^{2/3} }{\Gamma(2/3)}y^{1/3}{\rm e}^{-1/(9y)}  \;. \label{Hsmall}
\end{equation}

Using equation (\ref{cumul_lt.3}),
we can now express the probability density 
${\rm Prob.}\left[M(t)=L\right]$ of the maximum of the random walk in
a scaling form:
\begin{equation}
{\rm Prob.}\left[M(t)=L\right] = - \frac{ {\rm d}}{{\rm d}L} 
{\rm Prob.}\left[M(t)\ge L\right]= \frac{1}{(3t)^{1/3}}\,
g\left(\frac{L}{(3t)^{1/3}}\right)\;,
  \label{PMLg}
\end{equation}
where the scaling function $g(z)$ is simply related to the
scaling function $Y(y)$ 
and we deduce that
\begin{equation}
  g(z) =  z^{-4} \left. Y'(y)\right|_{y=1/(3z^3)}\;.
\end{equation}
Using the asymptotic behavior of $Y(y)$, we can then obtain
the asymptotic tails of $g(z)$ as
\begin{eqnarray}
  g(z) &\sim& \frac{3^{1/3}}{\Gamma(2/3)}\,z\, {\rm e}^{-z^3/3} 
\quad\quad\quad\quad\quad\quad\quad\quad\,\, \mbox{for} \quad z\to \infty \label{glarge}\\
  g(z) &\sim& \frac{3^{2/3} }{\Gamma(1/3)}\left[ 1 -\frac{2. 3^{2/3} \Gamma^2(2/3)}{\Gamma^2(1/3)}  z\ldots \right]\quad \mbox{for}\quad z \ll 1\;.
\label{gsmall}
\end{eqnarray}

\mre{In Fig. (\ref{fig.gz})
  we plot the results  of numerical simulations  of the  scaling distribution of the maximum, $g(z)$.
  We first note that the distribution is non-monotonic with $g(z)$ initially decreasing to a local minimum then rising to a local maximum before decreasing again.
  Also in Fig. (\ref{fig.gz}) we plot the tails of $g(z)$, equations \eqref{glarge} and \eqref{gsmall} and we see quantitative agreement with the simulation results at low and high $z$. The large
  $z$  form \eqref{glarge}  has a maximum at $z=1$ which is approached by the maximum in the numerical data.}

 It is useful to compare equation \eqref{glarge} with equation \eqref{Gz_sol}. We see that for large $z$,
the scaling function of the position distribution \eqref{Gz_sol}  and  that of the
maximum \eqref{glarge} have the same asymptotic tails up to an overall factor $1/2$.
This is similar to what occurs for a simple random walk, although in that case the tails are Gaussian. The small $z$ behaviour \eqref{gsmall}  for the scaling function is a constant with a linear correction. The constant is consistent with the large time limit of the survival probability \eqref{Qsmallz}. The linear correction contrasts with the case of a simple random walk where the correction to the constant term is quadratic in the scaling variable. 

\section{Generating function approach}
In sections \ref{sec:scaling} - \ref{sec:max}, we adopted a scaling approach to obtain long-time asymptotic results for the sluggish random walk problem. We now illustrate how a generating function approach may be employed to find the exact solution for all times. We will see that the long time limit of the solution obtained using the generating function approach recovers the results of the scaling approach. For
the sake of brevity, we restrict ourselves to the computation of the survival probability.

Consider again $Q(k_0,t)$, the survival probability for a walker starting at $k_0$ in the presence of a sink at the origin $k=0$. $Q(k_0,t)$  satisfies the backward master equation \eqref{bfp.1}. We define a generating function with parameter $\lambda$:
\begin{equation}\label{eq:G}
  {\cal G}(k_0,\lambda) = \sum_{t=0}^\infty \lambda^t Q(k_0,t)\;.
 \end{equation}
  Substituting \eqref{eq:G} into \eqref{bfp.1} and imposing the initial condition
   $ Q(k_0,0) = 1$,
 we obtain
\begin{equation}
    \left[ k_0 \frac{1-\lambda}{\lambda} + \frac{2}{\lambda}\right]
    {\cal G}(k_0,\lambda) -  {\cal G}(k_0+1,\lambda) - {\cal G}(k_0-1,\lambda)= \frac{k_0+2}{\lambda}\;.
\label{GF}
\end{equation}

\mre{We now compare the homogeneous part of  \eqref{GF} 
to the  recursion relation satisfied by Bessel functions of order $\mu$,
\begin{equation}
  J_{\mu-1}(x) + J_{\mu+1}(x) = \frac{2 \mu}{x} J_\mu(x)\;.
\end{equation}
Identifying
\begin{equation}
  x= \frac{2\lambda}{(1-\lambda)} \quad\mbox{and}\quad \mu= k_0+ \frac{2}{(1-\lambda)}\;, \label{xmu}
 \end{equation}
 one deduces that the homogeneous version of  \eqref{GF} (i.e. equating the lhs to zero) has Bessel functions as solution: 
    \begin{equation}
      {\cal G}_{hom}(k_0,\lambda) = B J_{\mu}(x)
    \end{equation}
    where $x$ and $\mu$ are defined in terms of  $k_0$ and $\lambda$ in \eqref{xmu}
and $B$ is a constant.
We have discarded the second solution, which is a Bessel function of the second kind $Y_{\mu}(x)$,  as it diverges as $k_0 \to \infty$.
A particular solution to \eqref{GF}  is $1/(1-\lambda)$ and the general solution to \eqref{GF} is therefore
\begin{equation}
      {\cal G}(k_0,\lambda) = B J_{\mu}(x) + \frac{1}{1-\lambda}\;.
    \end{equation}
    The boundary condition is ${\cal G}(0,\lambda) = 0$, which fixes the constant $B$, and we obtain
    the solution to \eqref{GF} as
\begin{equation}
  {\cal G}(k_0,\lambda) = \frac{1}{1-\lambda} \left[ 1 - \frac{J_{k_0+2/(1-\lambda)}( 2 \lambda/(1-\lambda))}{J_{2/(1-\lambda)}(2 \lambda/(1-\lambda))}\right]\;.
\label{GFsol}
\end{equation}
Equation \eqref{GFsol} is an exact expression for the generating function ${\cal G}(k_0,\lambda)$ in which $k_0$ takes all integer values $|k_0| =0, 1, 2 \ldots$.
In order to obtain exact expressions for all  survival probabilities at times $t$ one would need to expand the generating function in powers of $\lambda$, which remains a challenge.}

\mre{
However, it is possible to extract the long time asymptotic behaviour in a straightforward manner
by considering the $\lambda \to 1$ limit of \eqref{GFsol}.
Defining
\begin{equation}
  \lambda = 1- \epsilon\;,
\end{equation}
we require the asymptotic expansion of Bessel functions for large order and argument.
The required expansion is \cite{Olver54,SNEM17}
\begin{equation}
  J_{2/\epsilon +k_0}(2/\epsilon -b) \sim  \frac{1}{3^{2/3}\Gamma(2/3)} \epsilon^{1/3}
  - \frac{(k_0+b+1)}{3^{1/3}\Gamma(1/3)} \epsilon^{2/3} + \cdots
\end{equation}
for $\epsilon \to 0$.
Substituting this expansion with $b=2$ in \eqref{GFsol}, we find the leading behaviour  as $\lambda \to 1$, 
\begin{equation}
 {\cal  G}(k_0,\lambda) \simeq
  3^{1/3} \frac{\Gamma(2/3)}{\Gamma(1/3)}
 \frac{k_0}{(1-\lambda)^{2/3}}\;.
\end{equation}
Thus the leading singularity is at $\lambda^* =1$ and is  of the form $(\lambda^*-\lambda)^{-2/3}$.
We invoke the usual Tauberian theorem \cite{Wilf},
which states that if a generating function $G(\lambda) = \sum_t Q(t) \lambda^t$
has singularity nearest the origin  $G(\lambda) \simeq g_\gamma(\lambda^*-\lambda)^{-\gamma}$,  then $Q(t)$
has large $t$ asymptotic behaviour
\begin{equation}
  Q(t) \sim \frac{g_\gamma}{\Gamma(\gamma)} \frac{t^{\gamma-1}}{(\lambda^*)^{t+\gamma}}\;.
  \end{equation}
In our case we identify $\lambda^* =1$ and $\gamma = 2/3$ which implies  the
following large $t$ asymptotic behaviour for the survival probability starting from initial position $k_0$:
\begin{equation}
\label{Q_small.1}
  Q(k_0,t) \sim
  \frac{3^{1/3}}{\Gamma(1/3)}k_0 t^{-1/3}\;.
\end{equation}
In this expression $k_0$ can take any fixed value and $t \to \infty$.
This matches perfectly with the small $z$ 
asymptotic of the scaling behaviour in (\ref{scaling_final})
upon using the small $z$ expansion of $f(z)$ in equation (\ref{z0}).}

\section{Generalisation to the case where $\alpha \neq 1$}\label{sec:alphannot1}
Up to now, we have considered only the case where the probability of hopping to the right or left is proportional to $1/(|k| +2)$, i.e. the exponent $\alpha=1$ in the general expression for the hopping probability, $A(|k| +2)^{-\alpha}$. We now generalise to the case where $\alpha \neq 1$, i.e. the hopping probability is proportional to $1/(|k| +2)^\alpha$. The  scaling argument given in section 2 for the case $\alpha =1$ easily extends to general $\alpha$:
the typical number of steps taken after time $t$ is now $N \sim t/|k_{\rm typ}|^\alpha$ and the random walk scaling $ |k_{\rm typ}| \sim N^{1/2}$ then implies $|k_{\rm typ}| \sim t^{1/(2+\alpha)}$. The case $\alpha =1 $ recovers the
$t^{1/3}$ scaling of the sluggish random walk studied in the majority of this paper, and the case $\alpha =0$ recovers usual random walk scaling.

In the $\alpha \neq 1$ case,
equation  \eqref{ffp_scaling.1} generalises  for $k\geq 0$ to
\mre{
  \begin{eqnarray}
    \frac{\partial}{\partial t} P(k,t) &\approx&  \frac{\partial^2}{\partial k^2}\left[ \frac{1}{k^{\alpha}} P(k,t) \right] \label{FPgen}\\[1ex]
 &=& \frac{1}{k^\alpha}\left[ \frac{\partial^2}{\partial k^2} P(k,t) - \frac{2\alpha}{k}\, \frac{\partial}{\partial k} P(k,t)
+\frac{\alpha(\alpha+1)}{k^2}\, P(k,t)\right]\, .
\label{ffp_scaling.2}
  \end{eqnarray}
  }
Equation \eqref{ffp_scaling.2} can be put into the standard  \mre{Smoluchowski} form \eqref{ffp3},
where now $D(k) = 1/k^\alpha$ and $U(k) = 1/k^\alpha$.
One can again solve \eqref{ffp_scaling.2} by the scaling approach discussed earlier. For general positive $\alpha$ it is easy to show
that, as expected, the scaling variable  becomes $k/t^\nu$ where $\nu = (2+\alpha)^{-1}$.
Therefore  the solution  of \eqref{ffp_scaling.2} for $P(k,t)$ has a scaling form
\begin{equation}
  P(k,t) = t^{-\frac{1}{\alpha +2}}\, G\left( k\, t^{-\frac{1}{\alpha +2}}\right)\, ,
\label{Pgen}
\end{equation}
where the scaling function $G(z)$ is symmetric and, for positive $z$, satisfies the nontrivial differential equation
\begin{equation}
  G''(z) + \left( \frac{z^{\alpha +1}}{\alpha+2} - \frac{2 \alpha}{z} \right) G'(z)
  + \left( \frac{ z^\alpha}{\alpha + 2} + \frac{\alpha(\alpha+1)}{z^2}\right)G(z)=0\, ,
  \label{Ggen}
\end{equation}
with boundary condition $G(z) \to 0$ as $z \to \infty$.
Remarkably, this equation admits the simple solution, satisfying the boundary condition,
\begin{equation}
  G(z) = A\, z^\alpha\, \exp \left(-\frac{z^{\alpha+2}}{(\alpha +2)^2}\right)\;,
\end{equation}
where the normalisation constant $A$ is given by
\begin{equation}
  A^{-1} = 2 (\alpha +2)^{\frac{\alpha}{\alpha+2}}\, \Gamma \left( \frac{\alpha +1}{\alpha+2}\right)\;.
\end{equation}
Using the symmetry $G(z) = G(-z)$, the full solution for all $z$ can be written as
\begin{equation}
  G(z) = A\, |z|^\alpha\, \exp \left(-\frac{|z|^{\alpha+2}}{(\alpha +2)^2}\right)\;.
  \label{posgen}
\end{equation}
When $\alpha =0$  we recover the standard Gaussian result for a simple random walk, while for $\alpha =1$ we recover the result \eqref{Gz_sol} upon rescaling $z \to 3^{1/3} z$.
We note that for any $\alpha >0$ there is a trough, i.e. a cusp singularity, at $z=0$. The trough at $z=0$ disappears only for the case of simple diffusion  ($\alpha =0$).

Similar scaling analyses can be performed for the survival probability as well as the distribution of the maximum site visited to the right. We do not repeat the analysis,
but  just note that the scaling implies that the
asymptotic  decay of the survival probability is $Q(t) \sim t^{-1/(\alpha +2)}$
and the maximum scales as $M(t) \sim t^{1/(\alpha +2)}$ .

\section{Conclusion}

In this paper we have studied a random walk with space-dependent transition probabilities. Our study was motivated by trap models of slow dynamics, but in contrast to most such models, our trap depths are not random but instead increase logarithmically with distance $k$ from the origin. The dynamics of a particle moving on the lattice of traps follows an inhomogeneous random walk which has symmetric transition probabilities that decrease  with $k$ as $1/k$. Thus the motion of a walker  slows down as it goes further and further away from the origin, a phenomenon that we term `sluggish dynamics'. The  sluggish dynamics causes the typical distance explored up to time $t$ to grow subdiffusively as $t^{1/3}$, in contrast  to the standard $t^{1/2}$ law for a simple random walk. 

We used a scaling approach, in which the scaling variable is  $k/t^{1/3}$, to compute long-time asymptotic results for various properties of this inhomogeneous random walk: the position distribution, the survival probability in the presence of a sink at the origin, the joint survival and position distribution, and the distribution of the maximum distance to the right. Interestingly, the position distribution has a trough (a  cusp singularity) at the origin and is bimodal, with two peaks located at $|k| = (3t)^{1/3}$. The contrasts with the usual Gaussian distribution for simple diffusion (which has a single maximum at $k=0$). The bimodal distribution and the $t^{1/3}$ scaling reflect the sluggish nature of the dynamics.
The survival probability shows an asymptotic decay $\sim t^{-1/3}$
at large time, which contrasts with the  $t^{-1/2}$ decay for a simple random walk. The fact that the survival probability decays to zero as $t \to \infty$ implies that the walk is recurrent in $d=1$, as is the simple random walk. The distribution of the maximum of the walk up to time $t$ has a nontrivial scaling function.

We further showed how a generating function approach can be used to find exact solutions for all times. Using this approach to compute the survival probability in the presence of a sink at the origin, we recover our scaling result in the long-time limit. Application of the same generating function approach to other observables should be a straightforward extension.

Finally, we generalised the model to cases where the transition probability decays as $1/|k|^\alpha$ with positive $\alpha$. Except for $\alpha =0$ (simple random walk), the position distribution always shows a  trough at the origin ($k=0$),  where it exhibits a singularity, behaving as $|k|^\alpha$.
Remarkably, the scaling function for the position distribution takes on a simple form (equation 
\eqref{posgen}) and there is always a trough at the origin with associated singularity $|z|^\alpha$ for $\alpha >0$.

It is worthwhile comparing the behaviour of our sluggish random walk model with that of the Gillis model outlined in the introduction. In the continuum limit the Gillis model becomes diffusion in a logarithmic potential \cite{DLBK11,OPRA20} and the corresponding Fokker-Planck equation reads
\begin{equation}
\frac{\partial }{\partial t} P(k,t) =\frac{\partial}{\partial k} \left [ \frac{\partial}{\partial k} P(k,t)+
   \left(\frac{\partial}{\partial k} U(k)\right) P(k,t)\right]\, ,
  \label{ffpg}
\end{equation}
where the potential $U(k) = 2 \epsilon \ln |k|$. The relevant case for us is $\epsilon <0$  whereby the potential is repulsive and the particle is pushed away from the origin.
In this Gillis case, the solution for the time-dependent position distribution has scaling form \cite{DLBK11,OPRA20}
\begin{equation}
  P(k,t) \to \frac{1}{t^{1/2}} G_{\rm Gill} \left( \frac{k}{t^{1/2}}\right)
\end{equation}
where the scaling function, $G_{\rm Gill}(z)$,  is given by
\begin{equation}
G_{\rm Gill}(z)= \frac{2^{\epsilon-1/2}}{\Gamma(1/2 - \epsilon)}\, |z|^{-2\epsilon}\, e^{-|z|^2/2}\, .
\label{Gz_g}
\end{equation}
This is to be compared with the scaling function $G(z)$ \eqref{Gz_sol} for the sluggish random walk model (where the scaling variable is $z=k/(3t)^{1/3}$).
As with  \eqref{Gz_sol}, the scaling function \eqref{Gz_g} is bimodal,  with peaks at $z = \pm (-2\epsilon)^{1/2}$,
and has a trough at the origin.  However,
the model exhibits diffusive scaling and is thus not sluggish.
The difference between the sluggish random walk and  diffusion in a logarithmic potential is evident when one compares the Fokker Planck equations \eqref{ffp3} and \eqref{ffpg}. The key difference is the space-dependent  diffusion constant $D(k) =1/k$
appearing in \eqref{ffpg}, along with the \mre{effective} potential $U(k)= 1/k$ . It is these  features that lead to a change of the scaling
variable  to $z = k/(3t)^{1/3}$ and consequent sluggish behaviour. 

\mre{
  It is also of interest to compare our results with other works that have studied space-dependent diffusion processes.
  In \cite{CCM13} the following Langevin equation for position $x$ was considered
\begin{equation}
  \frac{{\rm d} x}{{\rm d} t}= \sqrt{2 D(x)} \eta(t),
    \label{Langevin}
  \end{equation}
  where $\eta(t)$ is white noise and $D(x) = x^{-\alpha}$ is a space-dependent diffusivity.
  Using the Stratonovich prescription for \eqref{Langevin}   implies the
  Fokker Planck equation
  \begin{equation}
    \frac{\partial}{\partial t} P(x,t) = \frac{\partial}{\partial x}\left[ D(x)^{1/2}
      \frac{\partial}{\partial x} \left[ D(x)^{1/2} P(x,t) \right] \right]\label{FPstrat}
\end{equation}
 from which the position distribution was obtained \cite{CCM13,SCT22a,SCT22b}. The distribution exhibits a trough at the origin for $\alpha >0$.
 Interestingly, it  can be shown \cite{ST22} that using the It\^o prescription for \eqref{Langevin} yields the
 Fokker Planck equation \eqref{FPgen} that we have obtained as the continuum descrption of the sluggish random walk.}

The sluggish random walk model and its analysis are straightforward to generalise to higher dimensions and other observables. For example, it would interesting to study the return probabilities and recurrence/transience transition in a higher dimension for general $\alpha$.
It would also be of interest to study the time for the walker to traverse from one maximum of the position distribution to the other.
 More generally our study has shown that inhomogenous space-dependent random walks can exhibit surprising properties and it remains to explore the full range of such behaviour.

\vspace{2mm}

The authors thank Juraj Szavits-Nossan for helpful discussions.
AZ acknowledges support of the INSPIRE fellowship from DST India and the Physics Computing Facility lab at UCSD. RJA was supported by the European Research Council under consolidator grant 682237 EVOSTRUC and by the Excellence Cluster Balance of the Microverse (EXC 2051 - Project-ID 390713860) funded by the Deutsche Forschungsgemeinschaft (DFG).
MRE thanks LPTMS for the award of a CNRS Visiting Professorship, during which this work was written up. 
For the purpose of open access, the authors have applied a Creative Commons Attribution (CC BY) licence to any Author Accepted Manuscript version arising from this submission.

\end{document}